\documentclass[conference]{IEEEtran}
\IEEEoverridecommandlockouts

\usepackage{cite}
\usepackage{graphicx}
\usepackage{xcolor}
\usepackage[per-mode=repeated-symbol]{siunitx}
\usepackage{glossaries}
\usepackage[hidelinks, bookmarks=false]{hyperref}
\usepackage[noabbrev,capitalise]{cleveref}
\usepackage{booktabs}
\usepackage{array}
\usepackage[symbol]{footmisc}
\usepackage{threeparttable}
\usepackage{makecell}
\usepackage{arydshln}
\usepackage{tikz}
\usepackage{eso-pic}
\usepackage{hyphenat}
\usepackage{minted}
\usepackage{microtype}
\usepackage{amsmath,amssymb,amsfonts}
\usepackage{soul}
\usepackage{algorithmic}
\usepackage{enumitem}
\usepackage{multirow}
\usepackage{xspace}
\usepackage{pifont}
\usepackage{changepage}
\usepackage{multicol}
\usepackage{amssymb}
\usepackage{printlen}
\usepackage{newtxtext,newtxmath}
\usepackage{textcomp}
\def\BibTeX{{\rm B\kern-.05em{\sc i\kern-.025em b}\kern-.08em
    T\kern-.1667em\lower.7ex\hbox{E}\kern-.125emX}}

\usepackage[free-standing-units,per-mode=repeated-symbol]{siunitx}
\usepackage[allow-number-unit-breaks]{siunitx}
\DeclareSIUnit\comp{COMP}
\DeclareSIUnit\flop{FLOP}
\DeclareSIUnit\flops{FLOPS}
\DeclareSIUnit\bps{bps}
\DeclareSIUnit\Bps{Bps}
\DeclareSIUnit\gate{GE}
\DeclareSIUnit\op{OP}
\DeclareSIUnit\macu{MACU}
\DeclareSIUnit\ops{OPS}
\DeclareSIUnit\core{core}
\DeclareSIUnit\request{request}
\DeclareSIUnit\cycle{cycle}
\DeclareSIUnit\teraops{TOPS}
\DeclareSIUnit\ghz{GHz}
\DeclareSIUnit\mhz{MHz}
\DeclareSIUnit[number-unit-product = ]\percent{\%}

\definecolor{MidnightBlue}{HTML}{191970}
\definecolor{Mint}{HTML}{3EB889}
\definecolor{EnglishRed}{HTML}{A4515C}
\definecolor{SelectiveYellow}{HTML}{FFBA08}
\definecolor{CyanProcess}{HTML}{08B2E3}
\definecolor{OliveDrab7}{HTML}{4D4730}
\definecolor{Red}{HTML}{FF0000}
\colorlet{color1}{MidnightBlue}
\colorlet{color2}{Mint}
\colorlet{color3}{EnglishRed}
\colorlet{color4}{SelectiveYellow}
\colorlet{color5}{CyanProcess}
\colorlet{color6}{OliveDrab7}
\colorlet{colorAlert}{Red}
\definecolor{PulpGreen}{HTML}{168638}
\definecolor{PulpBlue}{HTML}{1269b0}
\definecolor{PulpRed}{HTML}{a8322c}
\definecolor{PulpYellow}{HTML}{f2c100}

\Crefname{equation}{Eq.}{Eqs.}
\Crefname{figure}{Fig.}{Figs.}
\Crefname{tabular}{Tab.}{Tabs.}

\makeatletter \newcommand{\AddSpaceIfAnonymous}{\@ifclasswith{acmart}{anonymous}{\vspace{10mm}}{}} \makeatother
\PackageWarning{TODO:}{#1!}

\newacronym{pe}{PE}{Processing Element}
\newacronym{rv}{RV}{RISC-V}
\newacronym{ai}{AI}{Artificial Intelligence}
\newacronym{ml}{ML}{Machine Learning}
\newacronym{cpu}{CPU}{Central Processing Unit}
\newacronym{asic}{ASIC}{Application Specific Integrated Circuit}
\newacronym[longplural={Systems-on-Chip}]{soc}{SoC}{System-on-Chip}
\newacronym{fpga}{FPGA}{Field Programmable Gate Array}
\newacronym{asip}{ASIP}{Application Specific Instruction Processor}
\newacronym{gpp}{GPP}{General Purpose Processor}
\newacronym{gp}{GP}{general-purpose}
\newacronym{gpgpu}{GP-GPU}{General Purpose Graphics Processing Unit}
\newacronym{gpu}{GPU}{Graphics Processing Unit}
\newacronym{sm}{SM}{Streaming Multiprocessor}
\newacronym{cuda}{CUDA}{Compute Unified Device Architecture}
\newacronym{mpi}{MPI}{Message Passing Interface}
\newacronym{cots}{COTS}{Commercial-Off-The-Shelf}
\newacronym{soa}{SoA}{state-of-the-art}
\newacronym{roi}{ROI}{Return on Investments}
\newacronym
[
  longplural={Core Complexes}
]
{cc}{CC}{Core Complex}

\newacronym{lte}{LTE}{Long Term Evolution}
\newacronym{nr}{NR}{New Radio}
\newacronym{4g}{4G}{4th Generation}
\newacronym{5g}{5G}{5th Generation}
\newacronym{b5g}{B5G}{Beyond-5G}
\newacronym{6g}{6G}{6th Generation}
\newacronym{urll}{URLL}{Ultra-Reliable Low-Latency}
\newacronym{mmtc}{mMTC}{massive Machine Type Communications}
\newacronym{embb}{eMBB}{enhanced Mobile Broadband}
\newacronym{3gpp}{3GPP}{3rd Generation Partnership Project}
\newacronym{oran}{O-RAN}{Open-RAN}
\newacronym{ran}{RAN}{Radio Access Networks}
\newacronym{cran}{C-RAN}{Cloud Radio Access Networks}
\newacronym{gnb}{gNB}{Next Generation Node B}
\newacronym{pusch}{PUSCH}{Physical Uplink Shared Channel}
\newacronym{sdr}{SDR}{Software Defined Radio}
\newacronym{phy}{PHY}{Physical Layer}
\newacronym{cu}{CU}{Centralized Unit}
\newacronym{du}{DU}{Distributed Unit}
\newacronym{ru}{RU}{Remote Unit}
\newacronym{ue}{UE}{User Equipment}
\newacronym{ofdm}{OFDM}{Orthogonal Frequency Division Multiplexing}
\newacronym{ofdma}{OFDMA}{Orthogonal Frequency Division Multiple Access}
\newacronym{bf}{BF}{Beam Forming}
\newacronym{mimo}{MIMO}{Multiple-Input, Multiple-Output}
\newacronym{che}{CHE}{Channel Estimation}
\newacronym{dmrs}{DMRS}{Demodulation Reference Symbol}
\newacronym{tti}{TTI}{Transition Time Interval}
\newacronym{sc}{SC}{sub-carrier}

\newacronym{add}{add}{Add}
\newacronym{mul}{mul}{Multiply}
\newacronym{mac}{MAC}{Multiply\&Accumulate}
\newacronym{pmac}{p.mac}{Post-increment Multiply-accumulate}
\newacronym{axpy}{AXPY}{A Times X Plus Y}
\newacronym{dotp}{DOTP}{Dot Product}
\newacronym{sdotp}{SDOTP}{Sum Dot Product}
\newacronym{matmul}{MatMul}{Matrix Multiplication}
\newacronym{gemm}{GEMM}{General Matrix Multiplication}
\newacronym{mvm}{MVM}{Matrix-Vector Multiplication}
\newacronym{cfft}{CFFT}{Complex Fast Fourier Transform}
\newacronym{sysinv}{SysInv}{Linear System Inversion}
\newacronym{choldec}{CholDec}{Cholesky Decomposition}
\newacronym{mmse}{MMSE}{Minimum Mean Squared Error}
\newacronym{conv2D}{Conv2D}{2D-Convolution}
\newacronym{dct}{DCT}{Direct Cosine Transform}

\newacronym{sram}{SRAM}{Static Random-Access Memory}
\newacronym{dram}{DRAM}{Dynamic Random-Access Memory}
\newacronym{spm}{SPM}{Scratchpad Memory}
\newacronym{tcdm}{TCDM}{Tightly Coupled Data Memory}
\newacronym{IDol}{I\$}{Instruction Cache}
\newacronym{dma}{DMA}{Direct Memory Access}
\newacronym{axi}{AXI}{Advanced eXtensible Interface}
\newacronym{noc}{NoC}{Nework on Chip}
\newacronym{csr}{CSR}{Control Status Register}
\newacronym{hbm}{HBM2E}{High Bandwidth Memory}

\newacronym{ipc}{IPC}{instructions-per-cycle}
\newacronym{wfi}{WFI}{wait-for-interrupt}
\newacronym{raw}{RAW}{read-after-write}
\newacronym{ins}{INS}{instruction}

\newacronym{fpu}{FPU}{Floating Point Unit}
\newacronym{fpss}{FP-SS}{Floating Point Sub-System}
\newacronym{ipu}{IPU}{Integer Processing Unit}
\newacronym{divsqrt}{DIVSQRT}{Division and Square-Root Unit}
\newacronym{lsu}{LSU}{Load Store Unit}
\newacronym{dsp}{DSP}{Digital Signal Processing}
\newacronym{qlr}{QLR}{Queue-Linked Register}

\newacronym{eda}{EDA}{Electronic Design Automation}
\newacronym{ge}{GE}{Gate Equivalent}
\newacronym{fo4}{FO4}{Fan-Out-of-4}
\newacronym{beol}{BEOL}{Back-End-of-Line}
\newacronym{pnr}{PnR}{Place and Route}
\newacronym{ppa}{PPA}{Power, Performance and Area}

\newacronym{numa}{NUMA}{Non-Uniform Memory Access}
\newacronym{fc}{FC}{Fully-Connected}
\newacronym{isa}{ISA}{Instruction Set Architecture}
\newacronym{simd}{SIMD}{Single Instruction Multiple Data}
\newacronym{spmd}{SPMD}{Single Program Multiple Data}
\newacronym{cdf}{CDF}{Cumulative Distribution Function}
\newacronym{api}{API}{Application Programmable Interface}
\newacronym{rtl}{RTL}{Register Transfer Level}
\newacronym{sfr}{SFR}{Synchronization Free Region}
\newacronym{dsl}{DSL}{Domain-Specific Language}

\newacronym{int}{INT}{integer}
\newacronym{fp}{FP}{floating-point}

\newcommand\copyrighttext{\footnotesize \textcopyright This work has been accepted by 51st IEEE European Solid-State Electronics Research Conference (ESSERC'2025) and submitted to the IEEE for possible publication. Copyright may be transferred without notice, after which this version may no longer be accessible.}
\newcommand\copyrightnotice{%
    \begin{tikzpicture}[remember picture,overlay]
        \node[anchor=south,yshift=10pt] at (current page.south) {\fbox{\parbox{\dimexpr\textwidth-\fboxsep-\fboxrule\relax}{\copyrighttext}}};
    \end{tikzpicture}%
}

\begin{document}

\title{A \SI{410}{\giga\flop\per\second}, \num{64} RISC-V Cores, \SI{204.8}{\giga\Bps} Shared-Memory Cluster in \SI{12}{\nano\meter} FinFET with Systolic Execution Support for Efficient B5G/6G AI-Enhanced O-RAN
}

\ifx\blind\undefined
\author{%
Yichao Zhang\textsuperscript{*}\quad
Marco Bertuletti\textsuperscript{*}\quad
Sergio Mazzola\textsuperscript{*}\quad
Samuel Riedel\textsuperscript{*}\quad
Luca Benini\textsuperscript{*\dag}\quad
\\
{\small
 \textsuperscript{*}IIS, ETH Z\"{u}rich\quad%
 \textsuperscript{\dag}DEI, University of Bologna%
}
\\
{\small\itshape%
\textsuperscript{*}\{yiczhang, mbertuletti, smazzola, sriedel, lbenini\}@iis.ee.ethz.ch %
}
}
\else
\author{\centering{\textit{Authors omitted for blind review.}}}
\fi

\maketitle

\begin{abstract}
We present HeartStream, a \num{64}-RV-core shared-L1-memory cluster (\SI{410}{\giga\flop\per\second} peak performance and \SI{204.8}{\giga\Bps} L1 bandwidth) for energy-efficient AI-enhanced O-RAN.
The cores and cluster architecture are customized for baseband processing, supporting complex (16-bit real\&imaginary) instructions: multiply\&accumulate, division\&square-root, SIMD instructions, and hardware-managed systolic queues, improving up to \num{1.89}$\times$ the energy efficiency of key baseband kernels.
At \SI{800}{\mega\hertz}@\SI{0.8}{\volt}, HeartStream delivers up to \SI{243}{\giga\flop\per\second} on complex-valued wireless workloads.
Furthermore, the cores also support efficient AI processing on received data at up to \SI{72}{\giga\op\per\second}.
HeartStream is fully compatible with base station power and processing latency limits: it achieves leading-edge software-defined PUSCH efficiency (\SI{49.6}{\giga\flop\per\second\per\watt}) and consumes just \SI{0.68}{\watt} (\SI{645}{\mega\hertz}@\SI{0.65}{\volt}), within the \SI{4}{\milli\second} end-to-end constraint for B5G/6G uplink.
\end{abstract}


\copyrightnotice

\section{Introduction}
\begin{figure}[t!]
\centering
    \includegraphics[width=\linewidth]{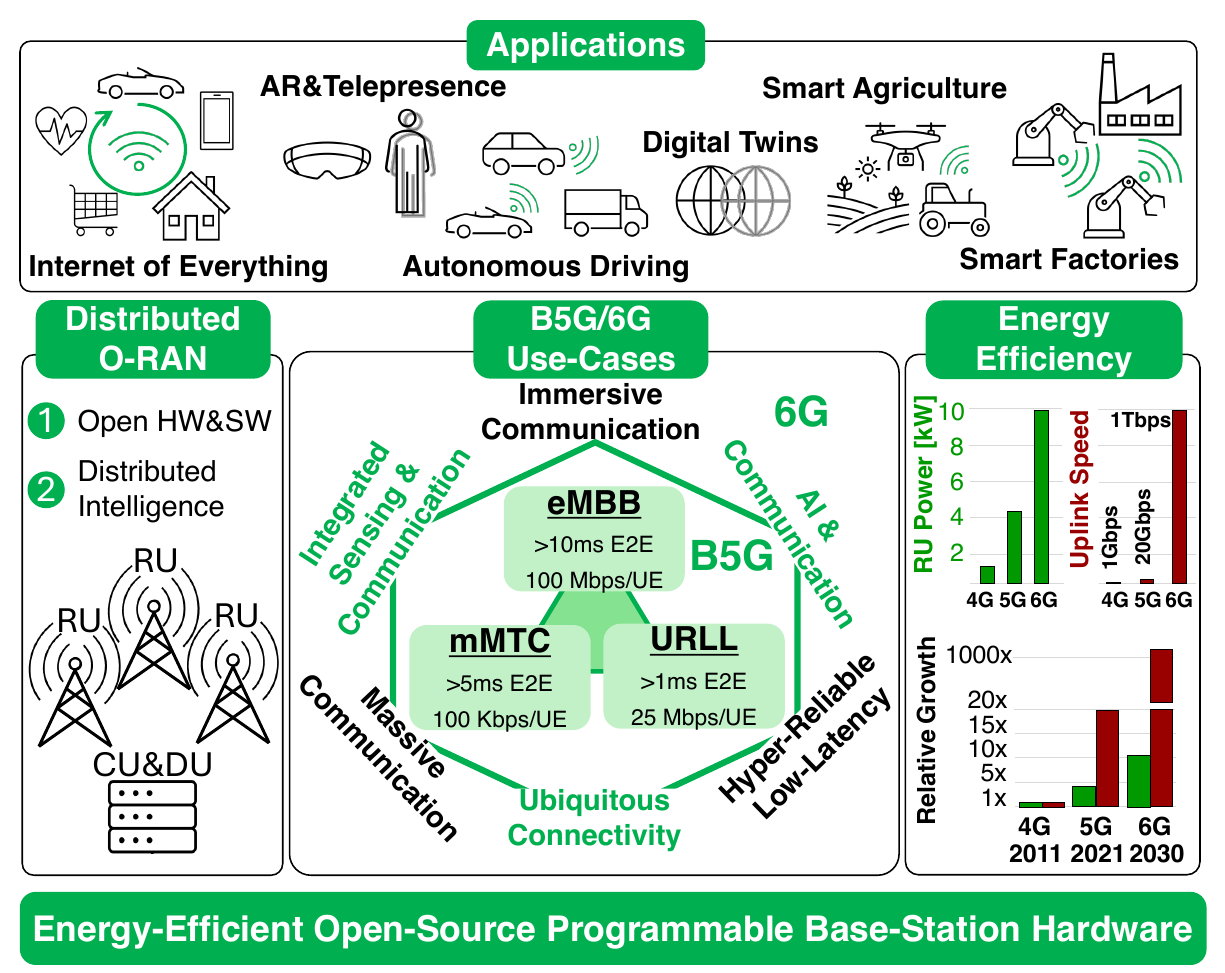}
    \vspace{-2em}
    \caption{Energy-efficient open-HW\&SW designs support B5G/6G O-RAN demanding use-case scenarios and a wide range of applications.}
    \label{fig:use_case}
\end{figure}

The evolution of 5G Cloud \gls{ran} toward 6G \gls{oran} (\Cref{fig:use_case}) relies on open programmable hardware\&software and distributed intelligence for the densification of network functions at the edge and the inter-operability of multi-vendor components~\cite{Gavrilovska_2020, Marinova_2024}.

\gls{b5g} and 6G require $>$\SI{20}{\giga\bps} uplink throughput, and $<$\SI{4}{\milli\second} end-to-end latency for high-end processing use-cases~\cite{Kumar_2024, ITU_2023}: \gls{urll}, \gls{mmtc}, and \gls{embb}.
This corresponds to an increase in the compute density of base stations, the most latency-critical processing components of the \gls{ran}.
Furthermore, the integration of \gls{ai} and communication will play a strategic role in the transition to 6G~\cite{aoudia2021end}, improving performance in complex deployment scenarios, but significantly increasing the computational demands on base stations.
Base station power is approaching thermal dissipation limits (\SI{10}{\kilo\watt}), while performance must meet a $50\times$ increase in uplink throughput by 2029~\cite{CL_2020, Ericsson_2021_2024}.
Thus, baseband processing requires ever-increasing energy efficiency at high performance at the \gls{ran} edge, coupled with programmability and flexibility to track fast-evolving heterogeneous workloads.

We present HeartStream, an open-source\footnote{\vspace{-2em}https://github.com/pulp-platform/mempool}, \gls{rv} cluster for efficient, software-defined \gls{oran}.
Our design demonstrates in silicon three innovations:
(A) a fully programmable \num{64}-core cluster, with \num{256}$\times1\text{-KiB}$ banks of shared-L1-memory (\SI{204.8}{\giga\Bps} bandwidth) through a scalable, hierarchical, low-latency (\num{1}-\num{5} cycles) interconnect;
(B) efficient integer, \gls{fp} (\num{32}/\num{16}/\num{8}-bit) and complex (\num{16}-bit real\&imaginary) enhanced cores, delivering \SI{410}{\giga\flop\per\second} (\SI{800}{\mega\hertz}@\SI{0.8}{\volt}) peak-performance;
and (C) interconnect and core extensions for systolic execution, boosting energy efficiency up to \SI{213}{\giga\flop\per\second\per\watt} in baseband and deep learning data-parallel workloads ($1.89\times$ over non-systolic kernel baseline).
HeartStream achieves up to \SI{8.99}{\giga\bps}@\SI{0.8}{\volt} for \gls{pusch} computing.
Even at the low-voltage, high-energy-efficiency corner (\SI{645}{\mega\hertz}@\SI{0.65}{\volt}), it meets the \SI{4}{\milli\second} end-to-end constraint for the baseband uplink of an 8x8 \gls{mimo} transmission, with \num{32} antennas, \num{8} beams, \num{8} users, and \SI{15}{\kilo\hertz} \gls{sc}-spacing on a \SI{15}{\mega\hertz}-FR1 band. 

\begin{figure*}[ht]
  \centering
  \includegraphics[width=\linewidth]{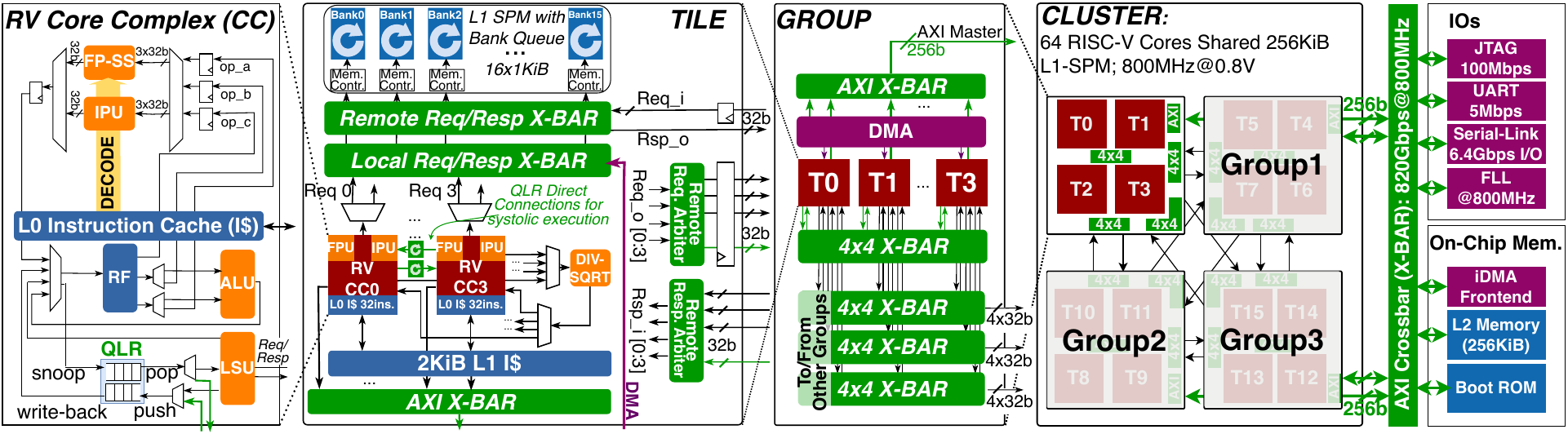}
  \vspace{-2em}
  \caption{HeartStream's 64 RISC-V cores shared-L1-memory hierarchical design architecture. L1 memory addresses are 32-bit interleaved across banks of 16 Tiles in 4 Groups. Each Tile's cores share an FP division/square-root unit. Core-Complex includes a 32b RISC-V core, IPU, FP-SS, and Systolic QLR.}
  \vspace{-1em}
  \label{fig:architecture}
\end{figure*}

\begin{figure}[ht]
\centering
    \vspace{-1em}
    \includegraphics[width=\linewidth]{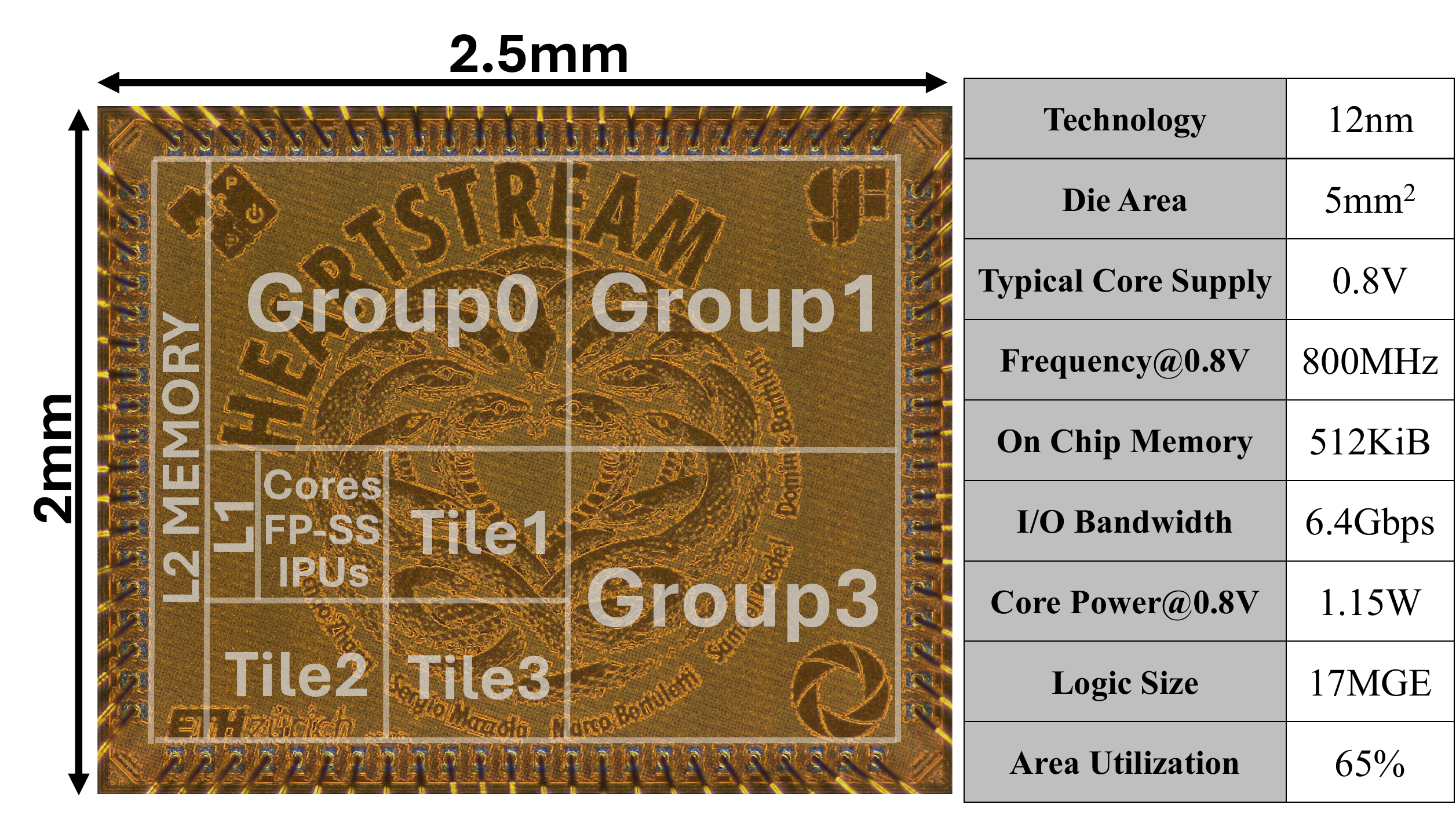}
    \vspace{-2em}
    \caption{Die micrograph and design summary. HeartStream was implemented in GlobalFoundries' \SI{12}{\nano\meter} FinFET technology on a \SI{5}{\milli\meter\squared} die. It achieves a \SI{65}{\percent} high utilization logic cell placement in the core area.}
    \vspace{-0.5em}
    \label{fig:die_shot}
\end{figure}

\begin{figure}[ht]
\centering
    \includegraphics[width=\linewidth]{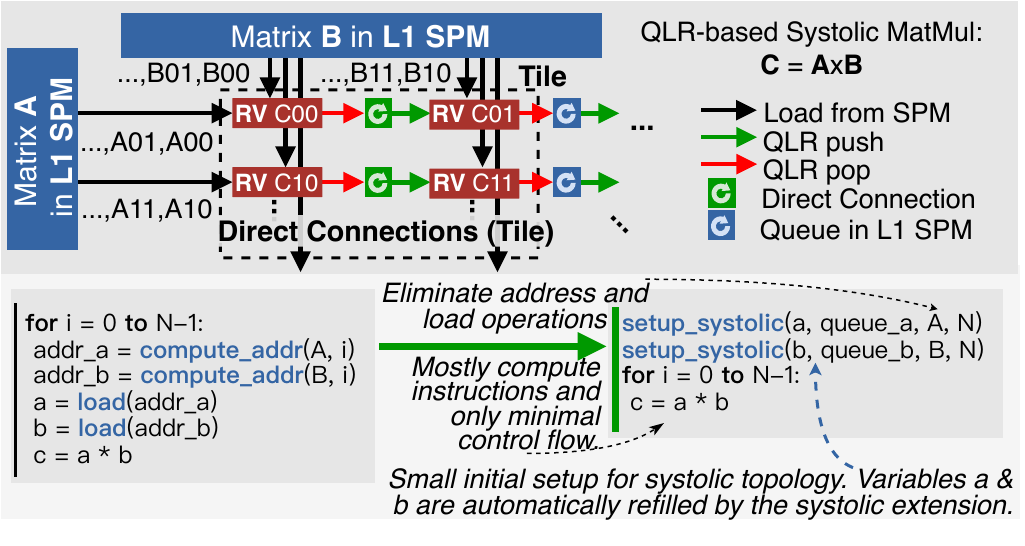}
    \vspace{-2em}
    \caption{In a systolic \acrshort{matmul} cores at the edge of the topology fetch from L1 and then forward data through QLR/memory queues, eliminating memory and control instructions; pseudocode shows that implicit inter-core communication eliminates many memory access and control instructions, boosting performance.}
    \label{fig:systolic_archi}
    \vspace{-1em}
\end{figure}

\section{Architecture}
HeartStream's shared-L1-memory architecture (\Cref{fig:architecture}) is inspired by MemPool~\cite{Riedel_2023}, aiming to reduce buffering, data movement, and synchronization overhead in transferring large baseband data chunks across memory hierarchies.
\Cref{fig:die_shot} presents the die shot and floorplan.
It features \num{64} cores connected to \num{256} 1-KiB L1-memory banks hierarchically (\num{4} \emph{Groups} of \num{4} \emph{Tiles} each), resulting in a \num{1}-\num{5} cycles low-latency access.
Each Tile contains \num{4} \glspl{cc} combining a single-stage latency-tolerant \num{32}-bit \gls{rv} \gls{gp} programmable core with an \gls{ipu} and a \gls{fpss}.
For efficient software-defined \gls{oran}, both units support domain-specific instructions: \gls{mac}, load-post-increment operations, SIMD operations, efficient complex arithmetic, widening sum-of-dot-product, and three-term addition instructions~\cite{Bertaccini_2024}.
One Tile-shared \gls{fp} division\&square-root unit helps to accelerate matrix-inversion for \gls{mimo} detection.
A key efficiency booster is hardware-supported flexible systolic execution with programmable topology~\cite{Mazzola_2024} within the shared-memory cluster.
Each core's \glspl{qlr} enable implicit inter-core register-file reads\&writes.
The \glspl{qlr} have direct connections within a Tile.
Across different Tiles, \gls{qlr} access requests are memory-mapped and routed by the cluster crossbar interconnects.
From the programming viewpoint, the \glspl{qlr} can be configured at the beginning of the program execution (pseudocode shown in~\Cref{fig:systolic_archi}).
After configuration, \gls{qlr} access is fully hardware-managed: data is automatically pushed/popped to/from the corresponding register-linked systolic streams.

The execution of a \gls{matmul} in systolic configuration is represented in~\Cref{fig:systolic_archi}.
In this topology, cores exchange the data of input matrices via \gls{qlr} connections and accumulate the outputs locally, reducing control-flow and memory-access overheads.
\gls{cfft} is also highly suitable for systolic execution.
We adopt a Cooley-Turkey decimation-in-time algorithm and map the butterfly stages to different groups of cores.
Cores pass each other the butterfly inputs/outputs without the need for a global inter-stage synchronization.
Additionally, twiddle-coefficients and bit-reversal addresses are assigned statically to a core, drastically reducing memory access.

Each Tile has a \num{256}-bit AXI port into a hierarchical interconnect to \SI{256}{\kibi\byte} of L2 memory, peripherals, and off-chip access.
A custom-designed DMA handles transfers between the physically-distributed L1 and L2 or off-chip memory.
The \num{16}-channel DDR Serial Link delivers \SI{6.4}{Gbps} throughput.

\section{Results}
\begin{figure}[t!]
\centering
    \includegraphics[width=0.93\linewidth]{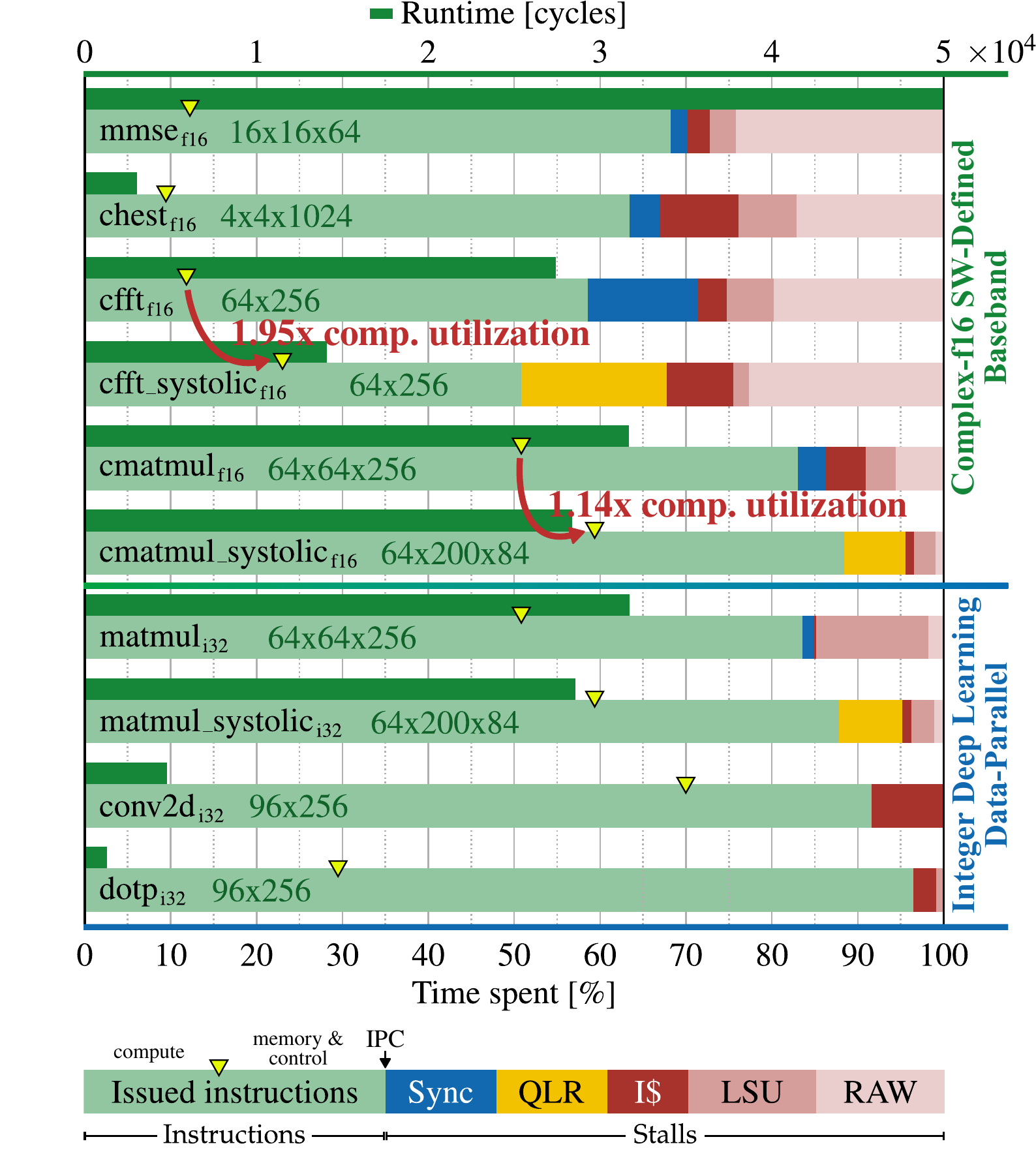}
    \caption{Absolute runtime and instruction/stall cycle fractions for 16-bit complex (real\&imaginary) baseband and 32-bit integer deep learning kernels. Systolic kernels achieve higher compute utilization and performance by reducing overhead instructions.}
    \label{fig:ipc}
    \vspace{-1em}
\end{figure}

HeartStream is designed for energy-efficient software-defined baseband processing.
Moreover, its fully programmable cores can also execute data-parallel \gls{ai} workloads on the received data streams.
This architectural flexibility plays a strategic role in enabling the convergence between wireless and \gls{ai} workloads planned by next-generation \gls{ran} systems.
Complex-valued baseband workloads achieve up to \SI{243}{\giga\flop\per\second}@\SI{0.8}{\volt} with \glspl{ipc} of \num{0.52}-\num{0.88} (\Cref{fig:ipc}).
On typical deep learning integer benchmarks (\gls{matmul}, \gls{conv2D}, and \gls{dotp}, with the largest input size fitting in the cluster L1 memory), HeartStream achieves \num{0.84}-\num{0.96} \gls{ipc} and up to \SI{72}{\giga\op\per\second}@\SI{0.8}{\volt}.

\begin{figure}[t!]
\centering
    \includegraphics[width=\linewidth]{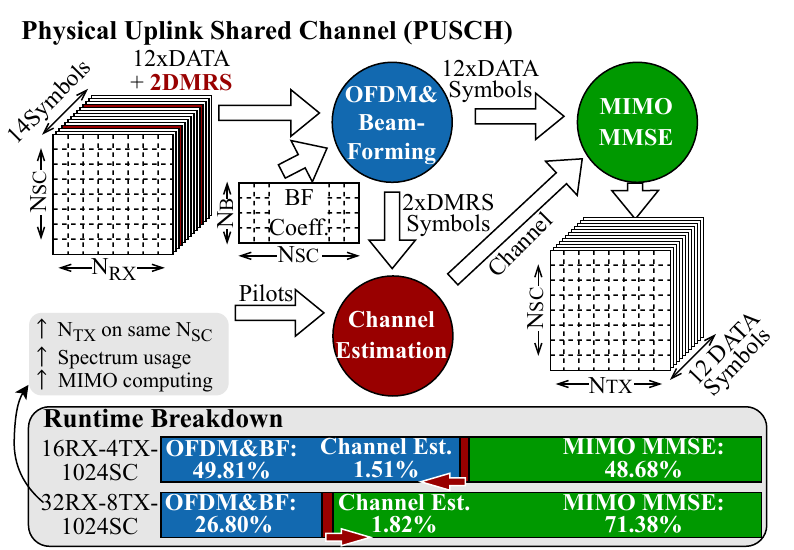}
    \vspace{-2em}
    \caption{The baseband PUSCH processing steps for a transition time interval with 14 symbols, 1024 Sub-Carries (SC) in 15kHz spacing, and compute runtimes breakdown for different $N_{RX}\times N_{TX}$ scenarios.}
    \vspace{-1em}
    \label{fig:pusch_pipeline}
\end{figure}

\begin{figure}[ht]
\centering
    \includegraphics[width=\linewidth]{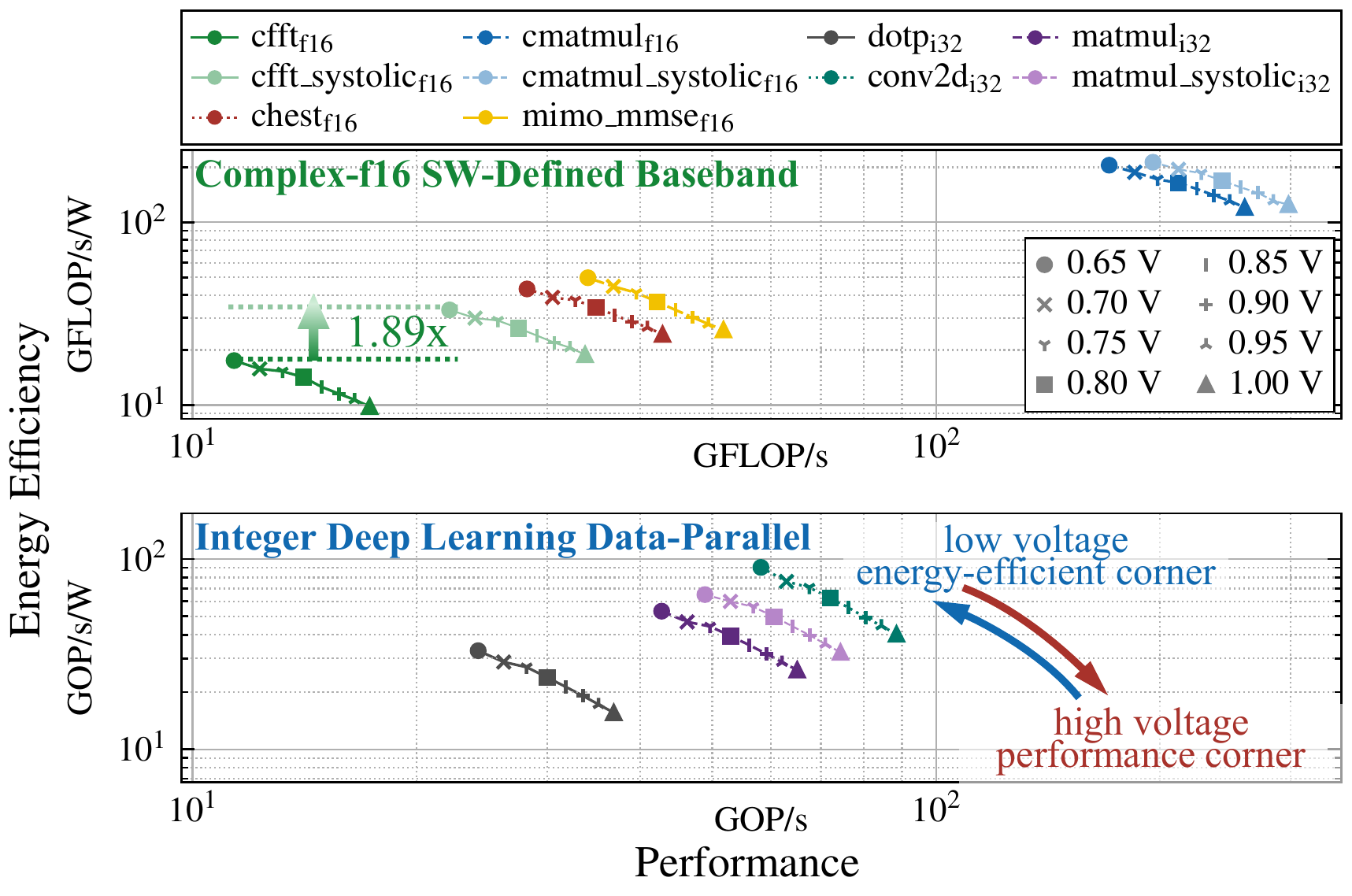}
    \vspace{-2em}
    \caption{HeartStream's efficiency and performance on key kernels for baseband and deep learning processing. The different core supply voltages target high energy efficiency or high performance. The systolic extension improves energy efficiency up to $1.89\times$.}
    \vspace{-1em}
    \label{fig:performance_efficiency}
\end{figure}

\gls{pusch} is one of the most compute- and time-critical channels of baseband processing. The uplink steps are represented in~\Cref{fig:pusch_pipeline}: in $<$\SI{4}{\milli\second} end-to-end latency, \num{14} symbols, each consisting of a $N_{RX}$ antennas $\times$ $ N_{SC}$ subcarriers matrix of complex numbers, undergo \gls{cfft} and C\gls{matmul} with known beamforming coefficients.
Two \gls{dmrs} symbols are used to estimate the transmission channel, and 12 data symbols are fed to the \gls{mmse} equalization, resulting in a detected complex number for each one of the $N_{TX}$ transmitters sending data over a subcarrier. 
In OFDM\&beamforming (\gls{cfft}\&C\gls{matmul}), systolic extensions allow \SI{50}{\percent} and \SI{12}{\percent} runtime reduction, respectively.
Pre-configuring a kernel-specific systolic computation reduces control and inter-core synchronization overhead in baseband and integer deep learning kernels.
The low-voltage operation (\SI{645}{\mega\hertz}@\SI{0.65}{\volt}) allows energy-efficient processing (\Cref{fig:performance_efficiency}): \SI{33.2}{\giga\flop\per\second\per\watt} OFDM and \SI{213}{\giga\flop\per\second\per\watt} beamforming, improving up to $1.89\times$ thanks to systolic extensions.
Even for $16\times16$ \gls{mimo}, energy-efficiency gains $+$\SI{13}{\giga\flop\per\second\per\watt}, with only $+$\SI{0.25}{\milli\second\per symbol} in runtime compared to \SI{800}{\mega\hertz}@\SI{0.8}{\volt}.

\begin{figure}[t!]
\centering
    \includegraphics[width=\linewidth]{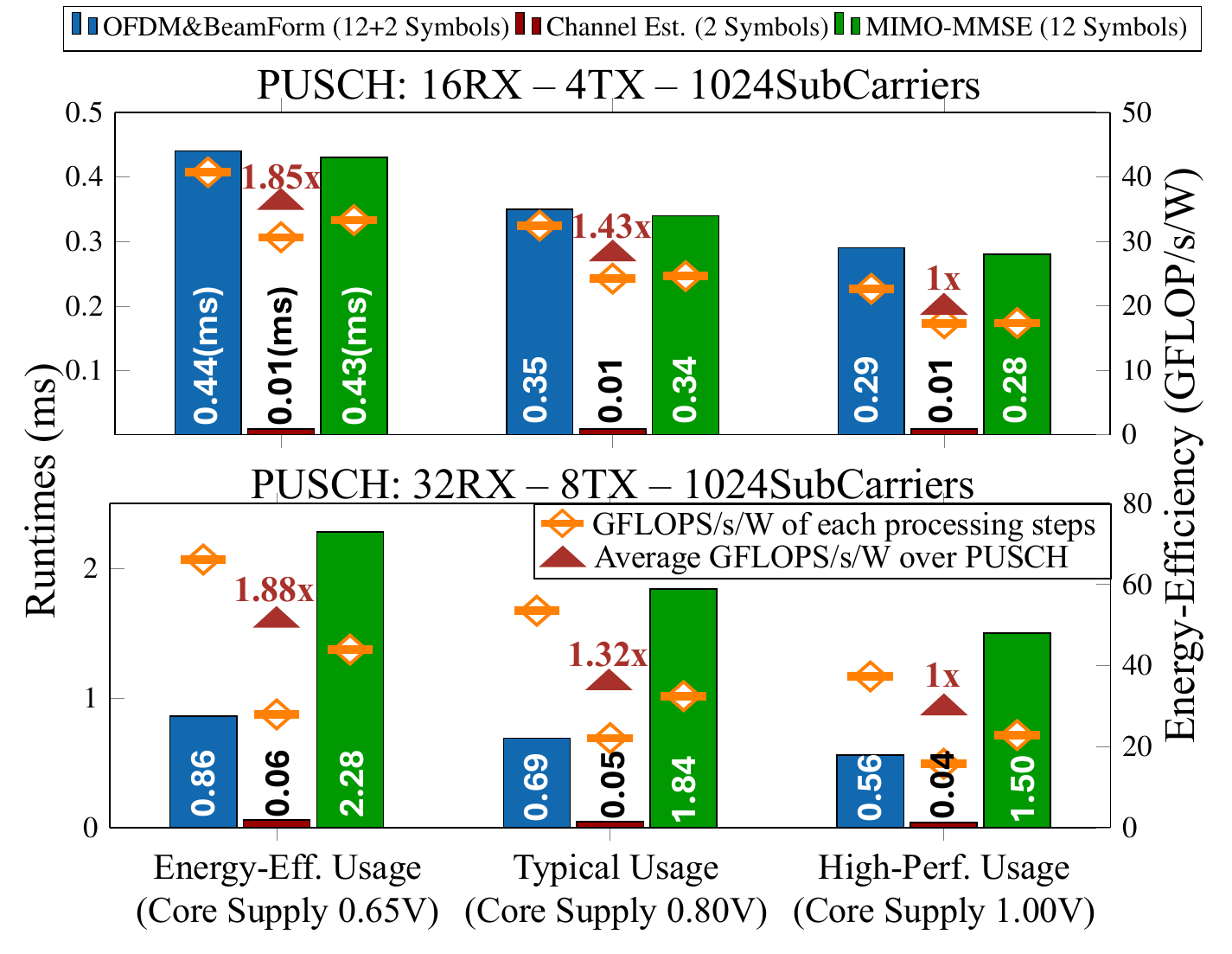}
    \vspace{-2em}
    \caption{Runtime and energy efficiency breakdown of PUSCH processing steps: low-voltage for low $N_{RX}\times N_{TX}$ achieves energy efficiency, and high-voltage for high $N_{RX}\times N_{TX}$ achieves high performance.}
    \vspace{-1em}
    \label{fig:pusch_results}
\end{figure}

\begin{figure}[t!]
\centering
    \includegraphics[width=0.95\linewidth]{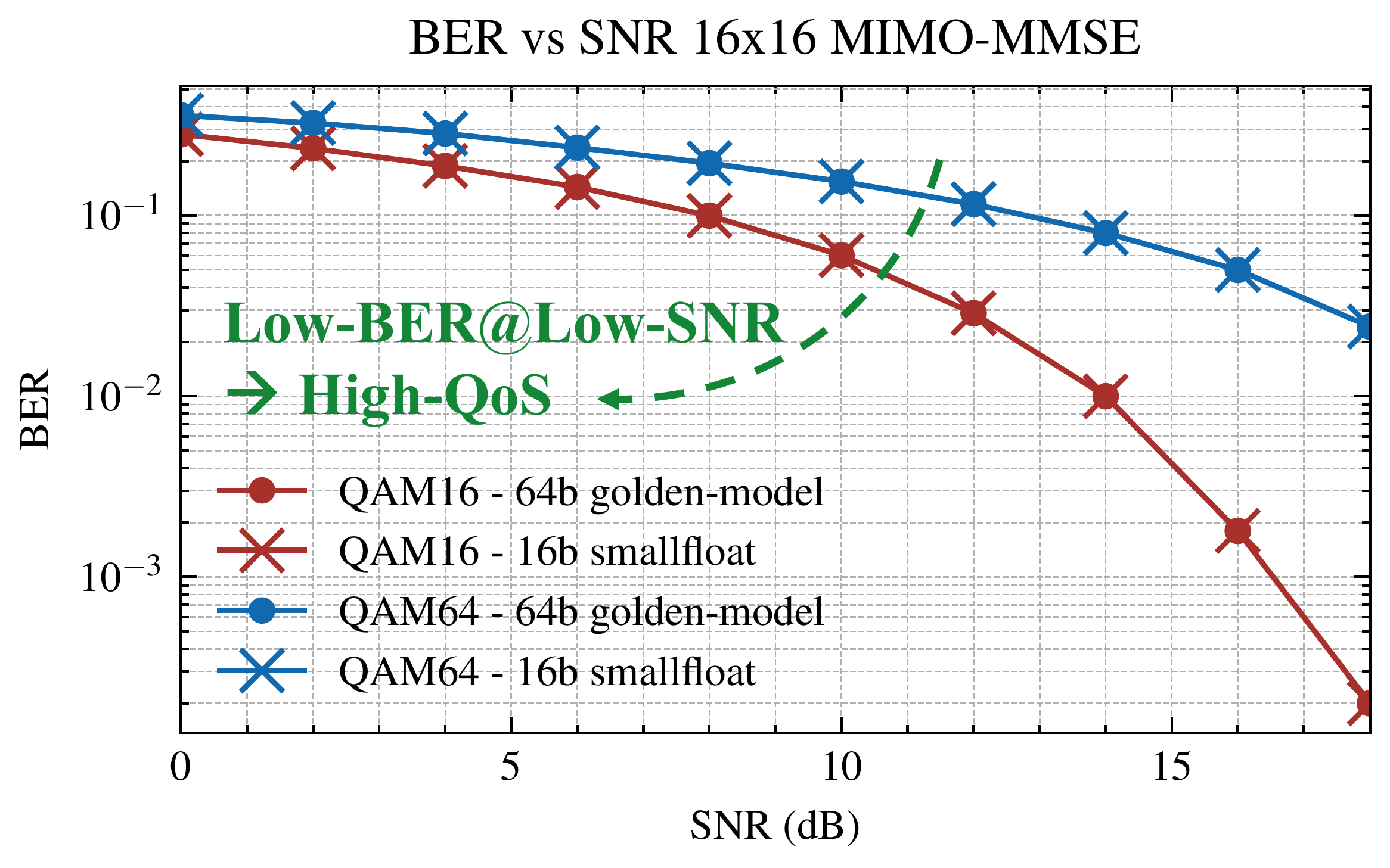}
    \vspace{-1em}
    \caption{BER vs. SNR of a 16x16 MIMO MMSE (AWGN channel), implemented with mixed-precision 16/32-bit floating-point extensions, yields the same results as the 64b golden model. Lower SNR at a given BER indicates higher Quality of Service (QoS).}
    \label{fig:ber_snr}
    \vspace{-1em}
\end{figure}

Two scenarios (\Cref{fig:pusch_results}) show different runtime distribution over \gls{pusch} steps: a $4\times4$ or $8\times8$ \gls{mimo} transmission, with $N_{RX}$=$16$-$32$ antennas, $N_B$=$4$-$8$ active beams, $N_{TX}$=$4$-$8$ transmitters, and \SI{15}{\kilo\hertz} \gls{sc}-spacing on a \SI{15}{\mega\hertz}-FR1 band.
HeartStream achieves an in-phase and quadrature antenna \gls{pusch} computing of up to \SI{8.99}{\giga\bps}@\SI{0.8}{\volt}.
In both scenarios, low-voltage operation increases energy efficiency up to \SI{49.6}{\giga\flop\per\second\per\watt} ($1.88\times$) and incurs only \SI{1.1}{\milli\second} slow-down in runtime, keeping the latency $<$\SI{4}{\milli\second} (\SI{3.2}{\milli\second}).
The compute budget available in the $4\times4$ \gls{mimo} can be invested in additional post-processing: all the presented deep learning benchmarks, for the largest problem size that fits in the L1, achieve high \num{37}-\SI{89}{\giga\op\per\second}@\SI{1}{\volt} and latencies far below the runtime constraints (\num{1}-\SI{32}{\micro\second}). 
Widening sum-of-dot-product of \emph{xsmallfloat} extensions is essential to keep complex arithmetic precision high in the matrix inversion operations implemented by \gls{mimo}-\gls{mmse}, yielding \SI{16.5}{dB} SNR at BER@$10^{-3}$dB (\Cref{fig:ber_snr}) for a $16\times16$ QAM16 \gls{mmse} problem.

\begin{table}[ht]
    \caption{Comparison with State-of-the-art Baseband Processing Design}
    \label{tab:soa}
    \centering
    \resizebox{\columnwidth}{!}{
        \setlength{\tabcolsep}{0pt} 
\begin{threeparttable}
\begin{tabular}{c|c|cccc|cc}
\hline
                                                                & \textbf{This Work$^a$}                                                                                       & \begin{tabular}[c]{@{}c@{}}ESSERC'24\\~\cite{Attari_2024}$^a$\end{tabular}                                 & \begin{tabular}[c]{@{}c@{}}CoolChip'22\\~\cite{Chen_2022}$^a$\end{tabular}                       & \begin{tabular}[c]{@{}c@{}}ISSCC'14\\~\cite{Noethen_2014}$^a$\end{tabular}                  & \begin{tabular}[c]{@{}c@{}}ESSCIRC'22\\~\cite{Oscar_2022}$^a$\end{tabular}           & \begin{tabular}[c]{@{}c@{}}ISSCC'24\\~\cite{Zhang_2024}$^b$\end{tabular}                 & \begin{tabular}[c]{@{}c@{}}VLSI'20\\~\cite{Wen_2020}$^b$\end{tabular}  \\ \hline
Technology                                                      & 12nm FinFET                                                                                              & 22nm FDX                                                                                 & 28nm                                                                             & 65nm                                                                     & 22nm FDX                                                           & 40nm                                                                    & 40nm                                                    \\
Freq.(MHz)                                                      & 800@0.8V                                                                                                 & 200-400@0.8V                                                                             & 800@0.9V                                                                         & 445@1.2V                                                                 & 293@0.8V                                                           & 200@1.1V                                                                & 290@1.1V                                                \\
I/O Gbps                                                        & 6.4                                                                                                      & -                                                                                        & -                                                                                & 10                                                                       & -                                                                  & 9.6                                                                     & 1.96                                                    \\ \hline
\begin{tabular}[c]{@{}c@{}}Processing\\ Element\end{tabular}    & {\color[HTML]{009900} \textbf{\begin{tabular}[c]{@{}c@{}}64 RISC-V\\ cores\end{tabular}}}                & \begin{tabular}[c]{@{}c@{}}1 RISC-V \&\\ 1 Vector,\\ 1 Systolic,\\ 1 Accel.\end{tabular} & \begin{tabular}[c]{@{}c@{}}1 RISC \&\\ 2 ARC,\\ 4 ASIPs,\\ 2 Accel.\end{tabular} & \begin{tabular}[c]{@{}c@{}}8 RISC \&,\\ 8 Vector,\\ 4 ASIPs\end{tabular} & \begin{tabular}[c]{@{}c@{}}16 PEs\\ Program.\\ ASIP.\end{tabular} & \begin{tabular}[c]{@{}c@{}}ASIC\\ Accel.\end{tabular}                   & \begin{tabular}[c]{@{}c@{}}ASIC\\ Accel.\end{tabular}   \\
\begin{tabular}[c]{@{}c@{}}Data\\ Precision\end{tabular}        & {\color[HTML]{009900} \textbf{\begin{tabular}[c]{@{}c@{}}Int32/16\\ FP 32/16/8\end{tabular}}}            & -                                                                                        & Int32/-, -                                                                       & \begin{tabular}[c]{@{}c@{}}Int16/-,\\ FP32/-\end{tabular}                & \begin{tabular}[c]{@{}c@{}}FP32/16/8\\ bfloat16\end{tabular}       & -                                                                       & -                                                       \\
\begin{tabular}[c]{@{}c@{}}GP-\\ Program.\end{tabular}          & {\color[HTML]{009900} \textbf{Yes}}                                                                      & {\color[HTML]{CB0000} No}                                                                & {\color[HTML]{CB0000} No}                                                        & {\color[HTML]{CB0000} No}                                                & {\color[HTML]{CB0000} No}                                          & {\color[HTML]{CB0000} No}                                               & {\color[HTML]{CB0000} No}                               \\
Execution                                                       & {\color[HTML]{009900} \textbf{\begin{tabular}[c]{@{}c@{}}SPMD,\\ SIMD,\\ Systolic\end{tabular}}}         & \begin{tabular}[c]{@{}c@{}}SIMD,\\ Systolic\end{tabular}                                 & \begin{tabular}[c]{@{}c@{}}VLIW,\\ SIMD\end{tabular}                             & \begin{tabular}[c]{@{}c@{}}SIMD,\\ ASIP-\\ MIMO\end{tabular}             & \begin{tabular}[c]{@{}c@{}}ASIP-\\ MIMO\end{tabular}                                                         & \textbf{-}                                                              & \textbf{-}                                              \\ \hline
\begin{tabular}[c]{@{}c@{}}Baseband\\ Processing\end{tabular}   & {\color[HTML]{009900} \textbf{\begin{tabular}[c]{@{}c@{}}Full B5G/6G\\ SW-Defined\\ O-RAN$^c$\end{tabular}}} & \begin{tabular}[c]{@{}c@{}}Partial RAN:\\ MIMO\\ Detector\end{tabular}                   & \begin{tabular}[c]{@{}c@{}}Partial RAN:\\ MIMO\\ Detector\end{tabular}           & \begin{tabular}[c]{@{}c@{}}Partial RAN:\\ LTE MIMO,\\ WiMAX\end{tabular} & \begin{tabular}[c]{@{}c@{}}MU-MIMO\\ Det./Dec.\end{tabular}        & \begin{tabular}[c]{@{}c@{}}MIMO\\ Det./Dec,\\ Channel Est.\end{tabular} & \begin{tabular}[c]{@{}c@{}}MIMO\\ Detector\end{tabular} \\
Peak Perf.                                                      & {\color[HTML]{009900} \textbf{\begin{tabular}[c]{@{}c@{}}410\\ GFLOP/s\end{tabular}}}                    & -                                                                                        & -                                                                                & \begin{tabular}[c]{@{}c@{}}3.6\\ GFLOP/s\end{tabular}                    & \begin{tabular}[c]{@{}c@{}}37.5\\ GFLOP/s\end{tabular}             & -                                                                       & -                                                       \\
\begin{tabular}[c]{@{}c@{}}MIMO\\ Workload\end{tabular}         & {\color[HTML]{009900} \textbf{Flex. Size$^d$}}                                                               & \begin{tabular}[c]{@{}c@{}}16x16\\ QAM64\end{tabular}                                    & \begin{tabular}[c]{@{}c@{}}4x4\\ QAM16\end{tabular}                              & \begin{tabular}[c]{@{}c@{}}4x4\\ QAM64\end{tabular}                      & {\color[HTML]{009900} \textbf{Flex. Size$^d$}}                         & \begin{tabular}[c]{@{}c@{}}8x8\\ QAM16\end{tabular}                     & \begin{tabular}[c]{@{}c@{}}256x32\\ QAM256\end{tabular} \\

\begin{tabular}[c]{@{}c@{}}PUSCH\\Comp. Gbps\end{tabular}      & \begin{tabular}[c]{@{}c@{}}8.99$^e$\end{tabular}                                                     & \begin{tabular}[c]{@{}c@{}}5.58$^f$\end{tabular}                                  & 7.30$^f$                                                                                & \begin{tabular}[c]{@{}c@{}}1.73$^f$\end{tabular}                   & \begin{tabular}[c]{@{}c@{}}1.76$^f$\end{tabular}                 & \begin{tabular}[c]{@{}c@{}}32.77$^f$\end{tabular}                     & \begin{tabular}[c]{@{}c@{}}26.13$^f$\end{tabular}       \\

\begin{tabular}[c]{@{}c@{}}PUSCH\\Comp.Gbps/W\end{tabular}      & \begin{tabular}[c]{@{}c@{}}8.26$^e$\end{tabular}                                                     & \begin{tabular}[c]{@{}c@{}}159.93$^f$\end{tabular}                                  & 7.18$^f$                                                                                & \begin{tabular}[c]{@{}c@{}}40.42$^f$\end{tabular}                   & \begin{tabular}[c]{@{}c@{}}33.26$^f$\end{tabular}                 & \begin{tabular}[c]{@{}c@{}}453$^f$\end{tabular}                     & \begin{tabular}[c]{@{}c@{}}1893$^f$\end{tabular}       \\

\begin{tabular}[c]{@{}c@{}}Deep-Learn.\\ Workloads\end{tabular} & {\color[HTML]{009900} \textbf{\begin{tabular}[c]{@{}c@{}}45.2$^g$\\ GOP/s/W\end{tabular}}}                   & -                                                                                        & -                                                                                & -                                                                        & -                                                                  & -                                                                       & -                                                       \\ \hline
\end{tabular}
\end{threeparttable}
    }
    \\ \scriptsize\raggedright\textsuperscript{a/b} Program./Non-Programmable solution.
    \scriptsize\raggedleft\textsuperscript{c} Open HW\&SW lower-PHY processing chain.
    \scriptsize\raggedright\textsuperscript{d} Software-Defined MIMO supports flexible No.TX/RX.
    \scriptsize\raggedright\textsuperscript{e} In-phase and quadrature antenna computing divided by runtime, 800MHz@0.8V.
    \scriptsize\raggedright\textsuperscript{f} MIMO processed by accelerator or specialized datapaths ASIP; technology normalized to 12nm and 0.8V core supply.
    \scriptsize\raggedright\textsuperscript{g} The energy efficiency of deep learning workload (Conv2D example).
\end{table}

In \Cref{tab:soa}, we compare HeartStream to \glspl{asic} \& \glspl{asip} for baseband processing.
HeartStream is the first open-source, \gls{rv}-based, and fully programmable \gls{oran} processor.
It delivers the highest peak performance (\si{\giga\flop\per\second}), and it achieves competitive throughput and energy efficiency compared to partially programmable designs with inflexible datapaths tailored for \gls{mimo} decoding~\cite{Attari_2024, Chen_2022, Noethen_2014, Oscar_2022}, while offering much greater flexibility to support multiple network scenarios with respect to fixed-function accelerators~\cite{Zhang_2024, Wen_2020}.
Further, it supports diverse \gls{mimo}-sizes, a full \gls{b5g}/6G uplink, and deep learning operators for the convergence between wireless and \gls{ai} processing in 6G \gls{ai}-native \glspl{ran}.

\section*{Acknowledgment}
\ifx\blind\undefined
     This work has received funding from the Swiss State Secretariat for Education, Research, and Innovation (SERI) under the SwissChips initiative.
\else
    \textit{Omitted for blind review.}
\fi

\Urlmuskip=0mu plus 1mu\relax
\def\UrlBreaks{\do\/\do-}
\bibliographystyle{IEEEtran}
\bibliography{bibliography}

\end{document}